\documentclass[smallcondensed]{svjour3}
\usepackage{balance}
\usepackage{amsmath,amssymb,amsfonts}
\usepackage{algorithmic}
\usepackage{graphicx}
\usepackage{textcomp}
\usepackage{url}
\usepackage{xcolor,colortbl}
\usepackage{todonotes}
\usepackage{framed}
\usepackage{siunitx}
\usepackage{multirow}
\usepackage{pgf-pie}
\usepackage{natbib}
\usepackage{graphicx} 
\makeatletter
\def\input@path{{.}{reports/}}
\makeatother
\graphicspath{{.}{reports/}} 
\newcommand{\rqn}[1]{\textbf{RQ{#1}:}}
\newcommand{\rquestion}[2]{\begin{framed}%
  \noindent\rqn{#1} #2
\end{framed}}

\newcommand{\rqtwo}{What is the prevalence of automated testing technologies in the FOSS mobile app development community?}
\newcommand{\rqthree}{Are today's mature FOSS Android apps using more automated testing than yesterday's?}
\newcommand{\rqfour}{How does automated testing relates to popularity metrics in FOSS Android apps?}
\newcommand{\rqfive}{How does automated testing affect code issues in FOSS Android apps?}
\newcommand{\rqsix}{What is the relationship between the adoption of CI/CD and automated testing?}

\newcommand{\hypothesis}[3]{%
  \item[] \textbf{#1}\\$H_0:$ #2.\\$H_1:$ #3.
}

\newcommand{\highlight}[1]{\begin{framed}\noindent\large \emph{#1} \end{framed}
}
\definecolor{darkgreen}{RGB}{40,160,40}
\begin{document}
\title{To the Attention of Mobile Software Developers: \\ Guess What, Test your App!}

\author{Luis Cruz \and Rui Abreu \and David Lo}
\institute{Luis Cruz
    \at INESC ID, Lisbon, Portugal\\\email{luiscruz@fe.up.pt}
    \and
    Rui Abreu
    \at INESC ID and IST, University of Lisbon, Lisbon, Portugal\\\email{rui@computer.org}
    \and
    David Lo
    \at School of Information Systems, Singapore Management University, Singapore\\\email{davidlo@smu.edu.sg}
}

\maketitle

\begin{abstract} Software testing is an important phase in the software
development lifecycle because it helps in identifying bugs in a software
system before it is shipped into the hand of its end users. There are numerous
studies on how developers test general-purpose software applications. The
idiosyncrasies of mobile software applications, however, set mobile apps apart
from general-purpose systems (e.g., desktop, stand-alone applications, web
services). This paper investigates working habits and challenges of mobile
software developers with respect to testing. A key finding of our exhaustive
study, using 1000 Android apps, demonstrates that mobile apps are still tested
in a very \emph{ad hoc} way, if tested at all. However, we show that, as in
other types of software, testing increases the quality of apps (demonstrated in
user ratings and number of code issues). Furthermore, we find evidence that tests are
essential when it comes to engaging the community to contribute to mobile open
source software. We discuss reasons and potential directions to address our
findings. Yet another relevant finding of our study is that Continuous
Integration and Continuous Deployment (CI/CD) pipelines are rare in the mobile
apps world (only 26\% of the apps are developed in projects employing CI/CD) -- we argue that one of the main reasons is
due to the lack of exhaustive and automatic testing.

\end{abstract}


\section{Introduction}
Over the last couple of years, mobile devices, such as smartphones and tablets,
have become extremely popular. According to a report by Gartner in
2015\footnote{Gartner's study on smartphone sales: \url{https://goo.gl/w757Vh}
(Visited on \today).}, worldwide sales of smartphones to end users had a record
2014 fourth quarter with an increase of 29.9\% from the fourth quarter of 2013,
reaching 367.5 million units. Amongst the reasons for the popularity of mobile
devices is the ever increasing number of mobile apps available, making
companies and their products more accessible to end users. As an example, nine
years after the release of the first smartphone running Android\footnote{The
first commercially available smartphone running Android was the HTC Dream, also
known as T-Mobile G1, announced on September 23, 2008:
\url{https://goo.gl/QPBdw9}}, there are 3.5 million mobile applications on
\emph{Google Play}\footnote{Google's market for Android
apps.}\textsuperscript{,}\footnote{Number of available applications in the
Google Play Store from December 2009 to December 2017 available at
\url{https://goo.gl/8P1KD7}.}.

Mobile app developers can resort to several tools, frameworks and services to
develop and ensure the quality of their apps~\citep{linares2017continuous}.
However, it is still a fact that errors creep into deployed software, which may
significantly decrease the reputation of developers and companies alike.
Software testing is an important phase in the software development lifecycle
because it helps in identifying bugs in the software system before it is
shipped into the hand of end users. There are numerous studies on how
developers test general-purpose software applications. The idiosyncrasies of
mobile software apps, however, set mobile apps apart from general-purpose
systems (e.g., desktop, stand-alone applications, web
services)~\citep{hu2011automating,picco2014software}.

Therefore, the onset of mobile apps came with a new ecosystem where traditional
testing tools do not always
apply~\citep{moran2017automated,wang2015mobile,maji2010characterizing}: complex
user interactions (e.g., swipe, pinch, etc.) need to be
supported~\citep{zaeem2014automated}; apps have to account for devices with
limited resources (e.g., limited power source, lower processing capability);
developers have to factor in an ever-increasing number of devices as well as OS
versions~\citep{khalid2014prioritizing}; apps typically follow a
weekly/bi-weekly time-based release strategy which creates critical time
constraints in testing tasks~\citep{nayebi2016release}. Moreover, manual
testing is not a cost-effective approach to assure software quality and ought
to be replaced by automated techniques~\citep{muccini2012software}.

This work studies the adoption of automated testing by the Android open source
community. We use the term ``automated testing'' as a synonym of ``test
automation'': the process in which testers write code/test scripts to automate
test execution. Automated Input Generation (AIG) techniques were not considered
in this study. We analyze the adoption of unit tests, Graphical User Interface
(GUI) tests, cloud based testing services, and Continuous Integration /
Continuous Deployment (CI/CD). Previous work, in a survey with 83 Android
developers, suggests that mobile developers are failing to adopt automated
testing techniques~\citep{kochhar2015understanding}. This is concerning since
testing is an important factor in software
maintainability~\citep{visser2016building}. We investigate this evidence by
systematically checking the codebase of 1000 Android projects released as Free
and Open Source Software (FOSS). Moreover, we delve into a broader set of
testing technologies and analyze the potential impact they can have in
different aspects of the mobile apps (e.g., popularity, issues, etc.).

As in related studies~\citep{krutz2015dataset}, we opted to use open source
mobile applications due to the availability of the data needed for our
analysis. Results cannot be extrapolated to industrial, paid apps, but we
provide. In particular, our work answers the following research questions:

%

\rquestion{1}{\rqtwo}

\noindent\textbf{Why and How:} It is widely accepted that tests play an
important role in assuring the quality of software code. However, the extent to
which tests are being adopted amongst the Android FOSS community is still not
known. We want to assess whether developers have been able to integrate tests
in their projects and which technologies have gained their acceptance. We do
that by developing a static analysis tool that collects data from an Android
project regarding its usage of test automation technologies. We apply the
tool to our dataset of $1000$ apps and analyze the pervasion of the different
technologies.

\vspace{1ex}
\noindent\textbf{Main findings:} FOSS mobile apps are still tested in a very
\emph{ad hoc} way, if tested at all. Testing technologies were absent in almost
60\% of projects in this study. \emph{JUnit} and \emph{Espresso} were the most
popular technologies in their category with an adoption of $36\%$ and $15\%$,
respectively. Novel testing and development techniques for mobile apps should
provide a simple integration with these two technologies to prevent
incompatibility issues and promote test code reuse.

\rquestion{2}{\rqthree}

\noindent\textbf{Why and How:} We want to understand how the community of
Android developers and researchers is changing in terms of adoption of
automated testing. In this study, we compare the pervasion of automated tests
in FOSS Android apps across different years.

\vspace{1ex}
\noindent\textbf{Main findings:}
Automated testing has become more popular in recent years. The trend shows that
developers are becoming more aware of the importance of automated testing. This
is particularly evident in unit testing, but GUI testing also shows a
promising gain in popularity.

\rquestion{3}{\rqfour}

\noindent\textbf{Why and How:} One of the goals of mobile developers is to
increase the popularity of their apps. Although many different things can
affect the popularity of apps, we study how it can be related to automated
tests. We run hypothesis tests over five popularity metrics to assess
significant differences between projects with and without tests.

\vspace{1ex}
\noindent\textbf{Main findings:} Tests are essential when it comes to engaging
the community to contribute to mobile open source software. We found that
projects using automated testing also reveal a higher number of contributors
and commits. The number of \emph{Github Forks}, \emph{Github Stars}, and ratings
from \emph{Google Play} users does not reveal any significant impact.

\rquestion{4}{\rqfive}

\noindent\textbf{Why and How:} The collection of code issues helps developers
assess whether their code follows good design architecture principles. It can
help developers avoid potential bugs, performance issues, or security
vulnerabilities in their software. We use the static analysis tool \emph{Sonar}
to collect code issues in our dataset of FOSS Android apps and study whether
automated testing brings significant differences.

\vspace{1ex}
\noindent\textbf{Main findings:} Automated testing is important to assure the
quality of software. This is also evident in terms of code issues. Projects
without tests have a significantly higher number of minor code issues.

\rquestion{5}{\rqsix}

\noindent\textbf{Why and How:} Previous work showed the adoption of CI/CD with
automated testing has beneficial results in software
projects~\cite{hilton2016usage,zhao2017impact}. For that reason, the adoption
of CI/CD is getting momentum in software projects. We want to study whether
CI/CD technologies have been able to successfully address the FOSS Android and
whether developers are getting the most out of CI/CD in their projects. We use
static analysis to collect data regarding the adoption of CI/CD technologies
and compare it to the adoption of automated testing. In addition, we discuss
how numbers differ from desktop software.

\vspace{1ex}
\noindent\textbf{Main findings:} CI/CD adoption in open source mobile app
development is not as predominant as in other platforms --- only 26\% of apps are
using it in their development process. We argue that one of the main reasons is
the lack of exhaustive and automatic testing --- results show evidence that
open source projects with CI/CD are more likely to automate tests.

%
%
%
%
%
%
%

In sum, our work makes the following contributions:

\begin{itemize}

\item We created a publicly available dataset with open source apps. The
dataset was built by combining data from multiple sources, including metrics of
source code quality, popularity, testing tools usage, and CI/CD services
adoption. Dataset is available
here: \url{https://github.com/luiscruz/android_test_inspector}.

\item We have studied the trends of the adoption of testing techniques in the
Android developer community and identified a set of apps that use automated
tests in their development cycle.

\item We have developed a tool for static detection of usage of state-of-art
testing frameworks. Available here:
\url{https://github.com/luiscruz/android_test_inspector}.

\item We have investigated the relationship of automated test adoption with quality and popularity metrics
for Android apps.

\item We have investigated the relationship between automated tests and CI/CD adoption.

\item We deliver a list of 54 apps that comply with testing best practices.

\end{itemize}

The remainder of this paper is organized as follows. Related work is discussed
in Section~\ref{sec:rw}. Section~\ref{sec:data_collection} outlines the
methodology used to collect data in our study. Following,
Sections~\ref{sec:rqtwo}--\ref{sec:rqsix} describe our methodology and present
and discuss the results for each proposed research question. In
Section~\ref{sec:hof}, we present a Hall of Fame with apps that comply with the
criteria of testing best practices. Threats to the validity are discussed in
Section~\ref{sec:t2v}. Finally, we draw our conclusions and point directions
for future work in Section~\ref{sec:conclusions}.

\section{Related Work}
\label{sec:rw}

Studies based on data collected from app stores have become a powerful source
of information with a direct impact on mobile software
teams~\citep{martin2017survey}. More works have contributed with datasets of
open source Android
apps~\citep{geiger2018graph,pascarella2018self,das2016quantitative}. Our paper
releases a dataset that differentiates by containing information regarding
testing practices in Android projects.

Previous work collected 627 apps from F-Droid to study the testing culture of
app developers~\citep{kochhar2015understanding}. It was found that at the time
of the analysis (2015) only $14\%$ of apps contained test cases and that only
$41$ apps had runnable test cases from which only 4 had line coverage above
$40\%$. In addition, the authors conducted a survey on $83$ Android app
developers and $127$ Windows app developers to understand the common testing
tools and the main challenges faced during testing. The most used framework was
\emph{JUnit}, being used by $18$ Android developers, followed by
\emph{Monkeyrunner} and \emph{Espresso}, with $8$ and $7$ developers,
respectively. According to developers in the survey, the main challenges while
testing are time constraints, compatibility issues, lack of exposure,
cumbersome tools, emphasis on development, lack of organization support,
unclear benefits, poor documentation, lack of experience, and steep learning
curve. Our work extends and completes the study by
\citeauthor{kochhar2015understanding} via a more extensive data sample (1000
Android apps) and additional analyses. We adopt a comprehensive
mining-software-repositories-cum-static-analysis approach to collect mobile
software code repositories and empirically assess the benefits of having tests,
rather than surveying developers. In addition, we compare the presence of tests
in the project with potential issues of the app, satisfaction level of end
users, among other popularity metrics. Moreover, we assess the use of
different testing tools using static analysis and provide insights
into observed trends on automated testing in the past years and compare the
testing culture with the adoption of CI/CD.

More works have attempted to capture the current picture of app testing.
\citeauthor{silva2016analysis} have studied $25$ open source Android apps in
terms of test frameworks being used and how developers are addressing
mobile-specific challenges~\citep{silva2016analysis}. Results show that apps
are not being properly tested, and tests for app executions under limited
resource constraints are practically absent. It suggests that a lack of
effective tools is one of the reasons for this phenomena. Our work
differentiates itself by considering a more representative sample of apps and
complements \citeauthor{silva2016analysis} by providing insights on how
developers and researchers can help bring new types of tests into the app
development community.

\citeauthor{coppola2017scripted} studied the fragility of GUI testing in
Android apps~\cite{coppola2017scripted}. The authors collected 18,930 open
source apps available on Github and analyzed the prevalence of five scripted
GUI testing technologies. However, toy apps or forks of real apps were not
factored out from the sample --- we understand that real apps were
underrepresented~\citep{cosentino2016findings,bird2009promises}. Thus, we
restrict our study to apps that were published in F-droid. In addition, we extend
our study to a broader set of testing technologies, while studying
relationships between automated testing and other metrics of a project.

\citeauthor{corral2015better} have compared the success of apps with quality
code metrics~\citep{corral2015better}. They analyzed a sample of $100$ apps and
consider a number of code metrics: \emph{Weighted Methods per Class},
\emph{Depth of Inheritance Tree}, \emph{Number of Children}, \emph{Response for
a Class}, \emph{Coupling between Objects}, \emph{Lack of Cohesion in Methods},
\emph{Cyclomatic Complexity}, and \emph{Logical Lines of Code}. Results
demonstrated that these metrics only have a marginal impact on the success of
the apps, showing that real drivers of user satisfaction are beyond source code
attributes. Given that mobile apps are very different from traditional
applications we find the above metrics too generic. We extend
\citeauthor{corral2015better}'s work by focusing on the impact of test
automation. Furthermore, besides user satisfaction, we also analyze a number of
code issues detected using static analysis and popularity metrics important for
the survival of an open source project (e.g., number of contributors).

Previous work has studied the state-of-the-art tools, frameworks, and services
for automated testing of mobile apps~\citep{linares2017continuous}. It revealed
that automated test tools should aid developers to overcome the following
challenges: 1) restricted time/budget for testing, 2) needs for diverse types
of testing (e.g., energy), and 3) pressure from users for continuous delivery.
Related work surveyed developers of open source apps to understand their main
barriers to mobile testing~\citep{linares2017developers}. Developers
identified easy maintenance, low overhead in the development cycle, and
expressiveness of test cases as important aspects that should be addressed by
existing frameworks.

Previous work has compared different techniques and tools for
AIG~\citep{choudhary2015automated,amalfitano2017general,zeng2016automated}.
\citeauthor{choudhary2015automated} have compared AIG testing tools in terms of
ease of use, ability to work on multiple platforms, code coverage, and ability
to detect faults~\citep{choudhary2015automated}. A follow-up study showed that
AIG techniques are not ready yet for an industrial setting since activity
coverage is dramatically low~\citep{zeng2016automated}. Our work does not scope
AIG techniques --- we focus on automated testing strategies that require the
creation of test cases. In addition, we differ by studying the prevalence of
testing tools and which test frameworks have actually gained the acceptance of
mobile developers.


Other works have empirically studied tests on open source
software.~\citeauthor{kochhar2013adoption} studied the correlation between the presence of test
cases and project development characteristics~\citep{kochhar2013adoption,kochhar2013empirical}. It
was found that tests increase the number of lines of code and the size of development teams. Our
work adds value to these contributions by providing insights in the context of mobile app
development, and by analyzing a broader set of metrics to study the potential benefits of automated
tests in mobile app development.

\citeauthor{hilton2016usage} analyzed $34,000$ open source projects on
\emph{GitHub} and surveyed $442$ developers~\citep{hilton2016usage} on the
implications of adopting CI/CD in open source software. Results showed that
most popular projects are using CI/CD and its adoption is continuously
increasing. A similar approach showed that developers are improving automated
tests after the adoption CI/CD~\citep{zhao2017impact}. Our work only focuses on
the relation between automated tests and CI/CD in the context of mobile
development, bringing some enlightenment on how the adoption of CI/CD differs
in mobile app development.


\section{Data collection}
\label{sec:data_collection}

\begin{figure}[htbp]
  \centering
    \includegraphics[width=0.8\columnwidth]{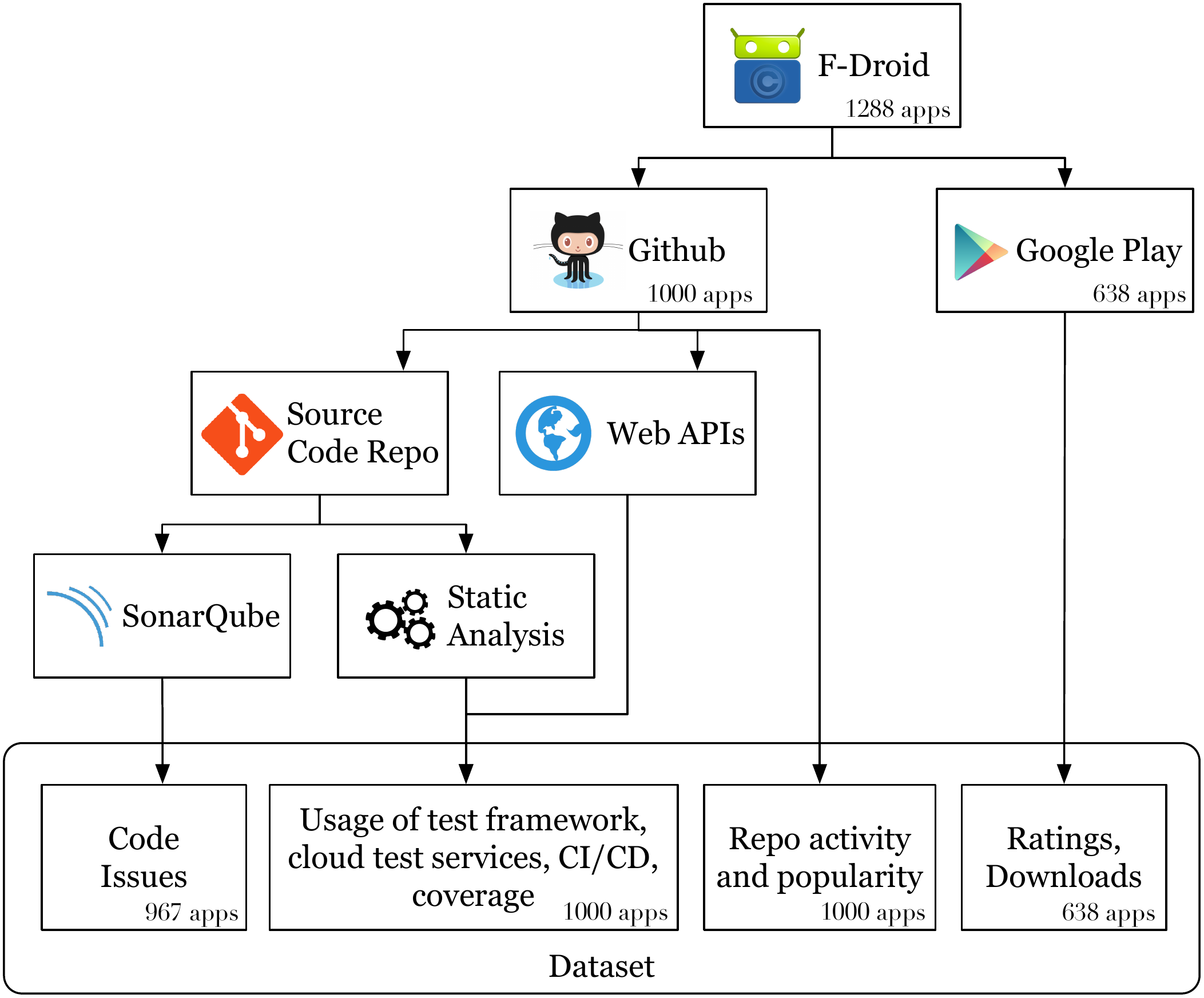}
  \caption{Flow of data collection in the study.}
  \label{fig:data_acquisition}
\end{figure}

Data was gathered from multiple sources, as presented in
Figure~\ref{fig:data_acquisition}. \emph{F-droid}, a catalog that lists
$2,800$ free and open source Android apps\footnote{F-droid's website:
\url{https://goo.gl/NPUusK} (Visited on \today). }, is used to obtain
metadata, package name, and source code repository. \emph{GitHub} is used to
collect activity and popularity metrics about the development of the app:
number of stars, number of contributors, number of commits, and number of
forks. Other popularity metrics are also gathered from \emph{Google Play
Store}: rating, and the number of users who rated the app. Test coverage
information is obtained from the cloud services \emph{Coveralls} and
\emph{Codecov}.

We extended the data by running the static analysis tool
\emph{Sonar}\footnote{Sonar's website: \url{https://goo.gl/svp88G} (Visited on
\today).} to collect quality-related metrics and potential bugs. We select
\emph{Sonar} because it integrates the results of the state-of-the-art analysis
tools \emph{FindBugs}, \emph{Checkstyle}, and \emph{PMD}. Furthermore, it has
been used with the same purpose in previous work~\citep{krutz2015dataset}.

For each project, we gather the total number of code issues detected by
\emph{Sonar}. We also count the number of code issues according to severity,
labeled as \emph{blocker} (issue with severe impact on app behavior and that
must be fixed immediately; e.g., memory leak), \emph{critical} (issue that might
lead to an unexpected behavior in production without impacting the integrity of
the whole application; e.g., badly caught exceptions), \emph{major} (issue that
might have a substantial impact on productivity; e.g., too complex methods), and
\emph{minor} (issue that might have a potential and minor impact on
productivity; e.g., naming conventions).

Directly comparing the number of issues in different projects can be
misleading: small projects are more likely to have fewer issues than large
projects, regardless of projects' code quality. To reduce this effect, we
controlled for the size of the project by normalizing the number of issues by the
number of files in a project, as follows:

\begin{equation}
I'(p) = \frac{I(p)}{F(p)},
\end{equation}
where $p$ is a given project, $I(p)$ the number of issues of $p$, and $F(p)$ the number of files.

Since one of the main goals in this work is to assess how apps are being
tested, we developed a tool to infer which testing frameworks a given project
is using\footnote{Source code repository of the tool created to inspect
automated testing technologies in Android projects:
\url{https://github.com/luiscruz/android_test_inspector}}. It works by fetching
the source code of the app and looking for imported packages and configuration
files. The efficacy of this tool was validated with a random sample of apps
which was manually labeled.

Table~\ref{tab:tools} lists all supported tools and frameworks aside with the
number of search results in \emph{StackOverflow}, as a proxy of popularity
among the developers' community. Unit test tools, user interface (UI)
automation frameworks, and cloud based testing services were selected based on
a previous survey on tools that support mobile
testing~\citep{linares2017continuous} and an online curated list of Android
tools\footnote{List of Android tools curated by Furiya:
\url{https://goo.gl/yLrWgW} (Visited on \today).}.

\begin{table}[htbp]
\caption{Android tools analyzed}
\begin{center}
\begin{tabular}{l r}
\hline
Name                    & StackOverflow Mentions$^{\mathrm{*}}$\\
\hline
\multicolumn{2}{c}{\emph{Unit testing}}\\
JUnit                   & $67,153$\\
AndroidJunitRunner      & $164$\\
RoboElectric            & $245$\\
RoboSpock               & $23$\\
\hline
\multicolumn{2}{c}{\emph{GUI testing}}\\
AndroidViewClient       & $474$ \\
Appium                  & $9,687$ \\
Calabash                & $1,856$ \\
Espresso                & $4,374$ \\
Monkeyrunner            & $1,299$ \\
PythonUIAutomator       & $0$ \\
Robotium                & $3,019$ \\
UIAutomator             & $1,918$\\
\hline
\multicolumn{2}{c}{\emph{Cloud testing services}}\\
Project Quantum         & $0$\\
Qmetry                  & $27$\\
Saucelabs               & $1,087$\\
Firebase                & $100,350$\\
Perfecto                & $224$\\
Bitbar\citep{kaasila2012testdroid} & $16$\\
\hline
\multicolumn{2}{c}{\emph{CI/CD} services}\\
Travis CI               & $3,662$\\
Circle CI                & $377$\\
AppVeyor                & $655$\\
CodeShip                & $564$\\
CodeFresh               & $6$\\
Wercker               & $200$\\
\hline
\multicolumn{2}{c}{$^{\mathrm{*}}$StackOverflow mentions as of January 26, 2018}\\
\hline
\end{tabular}
\label{tab:tools}
\end{center}
\end{table}

We also collect information about the usage of Continuous Integration and Continuous Delivery
(CI/CD) services in our study: \emph{Travis CI}, \emph{Circle CI}, \emph{AppVeyor},
\emph{Codeship}, \emph{Codefresh}, and \emph{Wercker}. The selection is based on CI/CD services
that have a free plan for open source projects and which adoption can be automatically assessed ---
i.e., either they save their configuration in the code repository or have an open API that can be
accessed with the \emph{GitHub} organization and project name. Self-hosted CI/CD platforms (e.g.,
\emph{GoCD}, \emph{Jenkins}) are not included in this list. Although this is a subset of CI/CD
services that can be used in a project, previous work found that \emph{Travis CI} and \emph{Circle
CI} have more than $90\%$ of share in \emph{GitHub} projects using CI/CD
services~\citep{hilton2016usage}.

We analyzed Android apps that are open source and published in \emph{F-droid}.
The most popular version control repository is \emph{GitHub}, being used by
around 80\% of projects. To make data collection clean, only projects using
\emph{GitHub} were considered. No other filtering was applied except in
particular analyses that required data that was not available for all apps
(e.g., \emph{Google Play}'s ratings).

\begin{figure}[htbp]
  \centering
    \includegraphics[width=0.9\columnwidth]{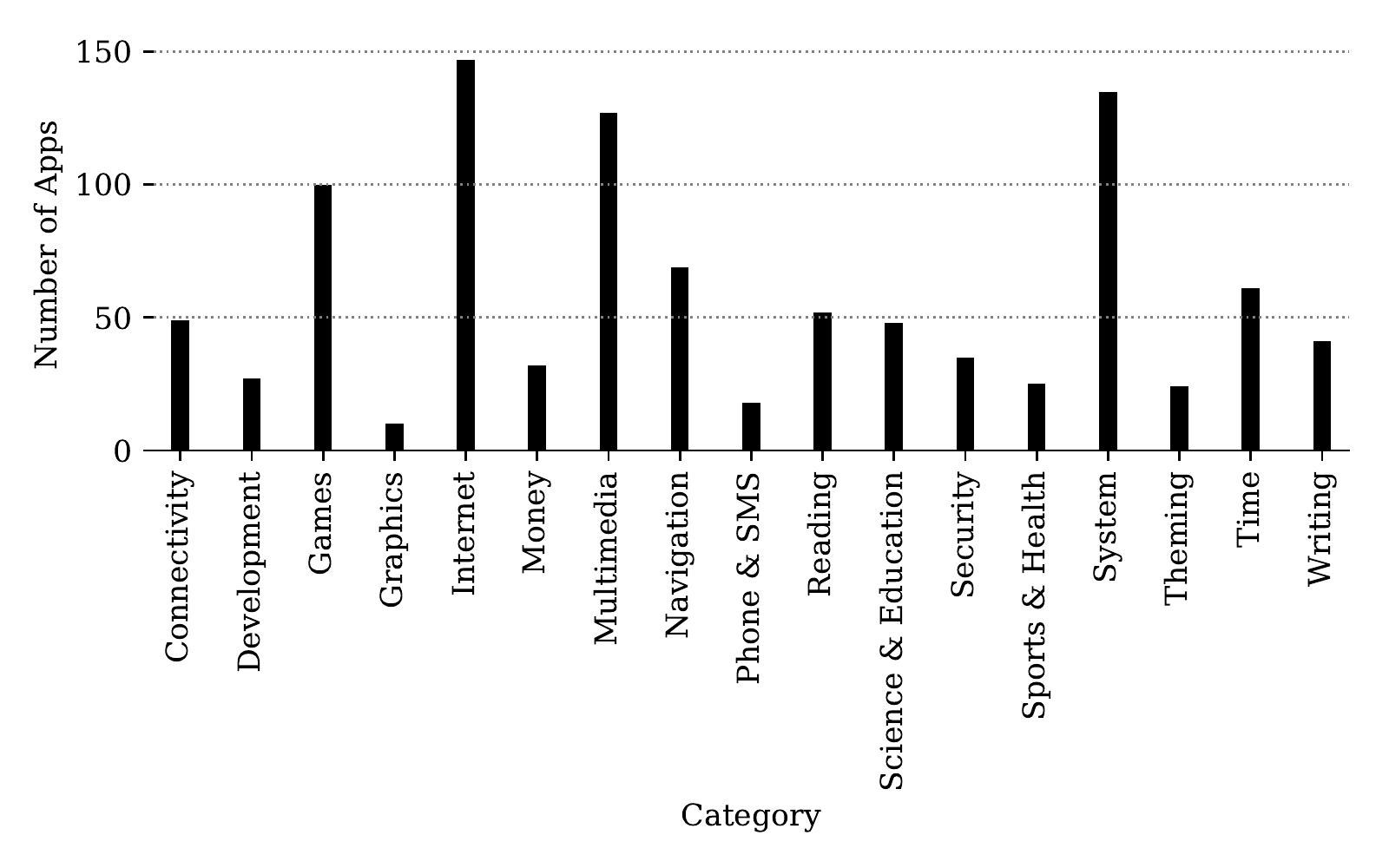}
  \caption{Categories of apps included in our study with the corresponding app count for each category.}
  \label{fig:categories}
\end{figure}

\begin{figure}[htbp]
  \centering
    \includegraphics[width=0.55\columnwidth]{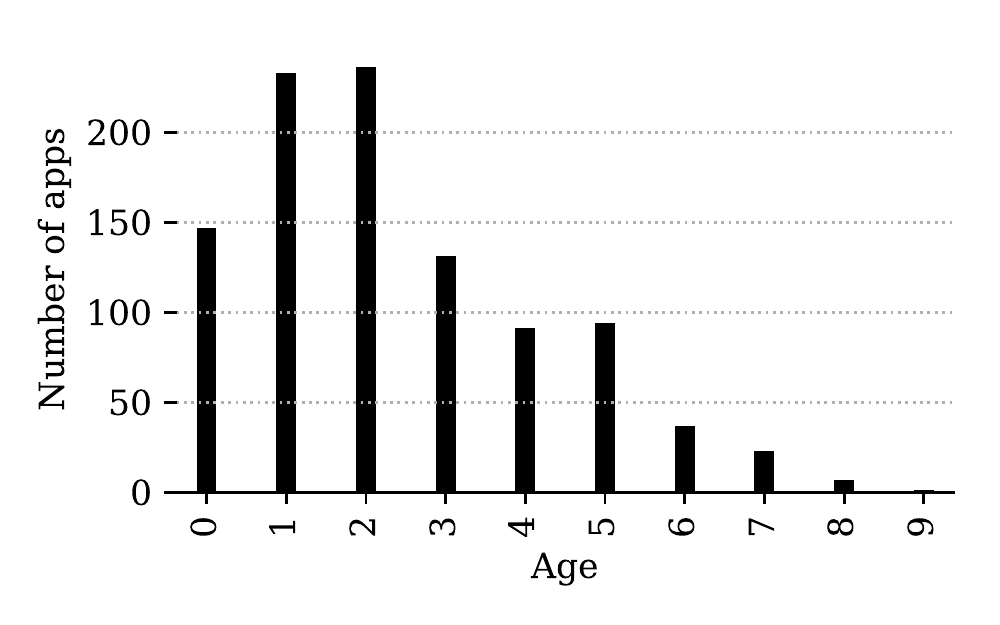}
  \caption{Distribution of apps by age.}
  \label{fig:apps_by_age}
\end{figure}

Although \emph{F-droid}'s documentation reports that it hosts a total $2,800$
apps\footnote{As reported in the \emph{F-droid}'s wiki page \emph{Repository
Maintenance}: \url{https://goo.gl/VfEQMg} (Visited on January 26, 2018).}, only
$1288$ actually make it to the end user catalog. As we restrict our study to
projects using \emph{GitHub}, in total we analyze $1000$ Android apps, roughly
$35GB$ of source code collected between September 1--8, 2017. Apps in the
dataset are spread amongst 17 categories, as presented in
Figure~\ref{fig:categories}, and are aged up to 9 years. The distribution of
apps by age is presented in Figure~\ref{fig:apps_by_age}.

Since in a few projects the static analysis tool \emph{Sonar} does not
successfully run, we collect code issues data for 967 apps, analyzing a total
of $329,676$ files. Additional data gathered from the \emph{Google Play} store
is available for 638 apps.

\subsubsection*{Reproducibility-oriented Summary}
\label{sec:reproducibility}

To power reproducibility, based on previous guidelines for app store
analyses~\citep{martin2017survey}, our work is best described as follows:

\begin{itemize}
    \item[] \textbf{App Stores used to gather collections of apps.} We use apps
    available on \emph{F-Droid} and combine it with data available on
    \emph{Google Play} store.

    \item[] \textbf{Total number of apps used.} The study comprises $1000$ apps.

    \item[] \textbf{Breakdown of free/paid apps used in the study.} Only free
    apps are listed in our dataset.

    \item[] \textbf{Categories used.} Apps in this study are spread across 17
    categories. The distribution of apps is illustrated with the bar chart of
    Figure~\ref{fig:categories}.

    \item[] \textbf{API usage.} We collect usage of APIs related to test
    automation exclusively.

    \item[] \textbf{Whether code was needed from apps.} Source code was required
    given the nature of analyses performed in the study.

    \item[] \textbf{Fraction of open source apps.} Open source apps are used
    exclusively.

    \item[] \textbf{Static analysis techniques.} We analyze source code with a
    self-developed tool for detection of tools, frameworks, and services' usage
    in the app's project and the static analysis techniques provided by
    \emph{SonarQube} to gather code issues.

\end{itemize}

All scripts and tools developed in this work are publicly available with an
open source license: \url{https://luiscruz.github.io/android_test_inspector/}. The same
applies to the whole dataset, for the sake of reproducibility.

%

\section{\rqtwo{} (RQ1)} 
\label{sec:rqtwo}


Testing is an essential task in software projects, and mobile apps are no
different. Given the specific requirements of mobile apps, conventional
approaches do not always apply. Thus, we want to assess how the FOSS mobile app
development community is addressing automated testing. In particular, we study
which testing approaches and technologies are most popular while discussing
potential factors.

We compare the frequency of the automated testing technologies employed in the
development of the apps in the dataset. The state-of-the-art technologies
listed earlier in Table~\ref{tab:tools} were included, dividing them into three
different categories: Unit testing, GUI testing, and Cloud testing services. We
resort to data visualizations and descriptive statistics to analyze the
frequency of technologies in the dataset.

\subsection{Results}

\begin{figure}[htbp]
  \centering
    \includegraphics[width=\columnwidth]{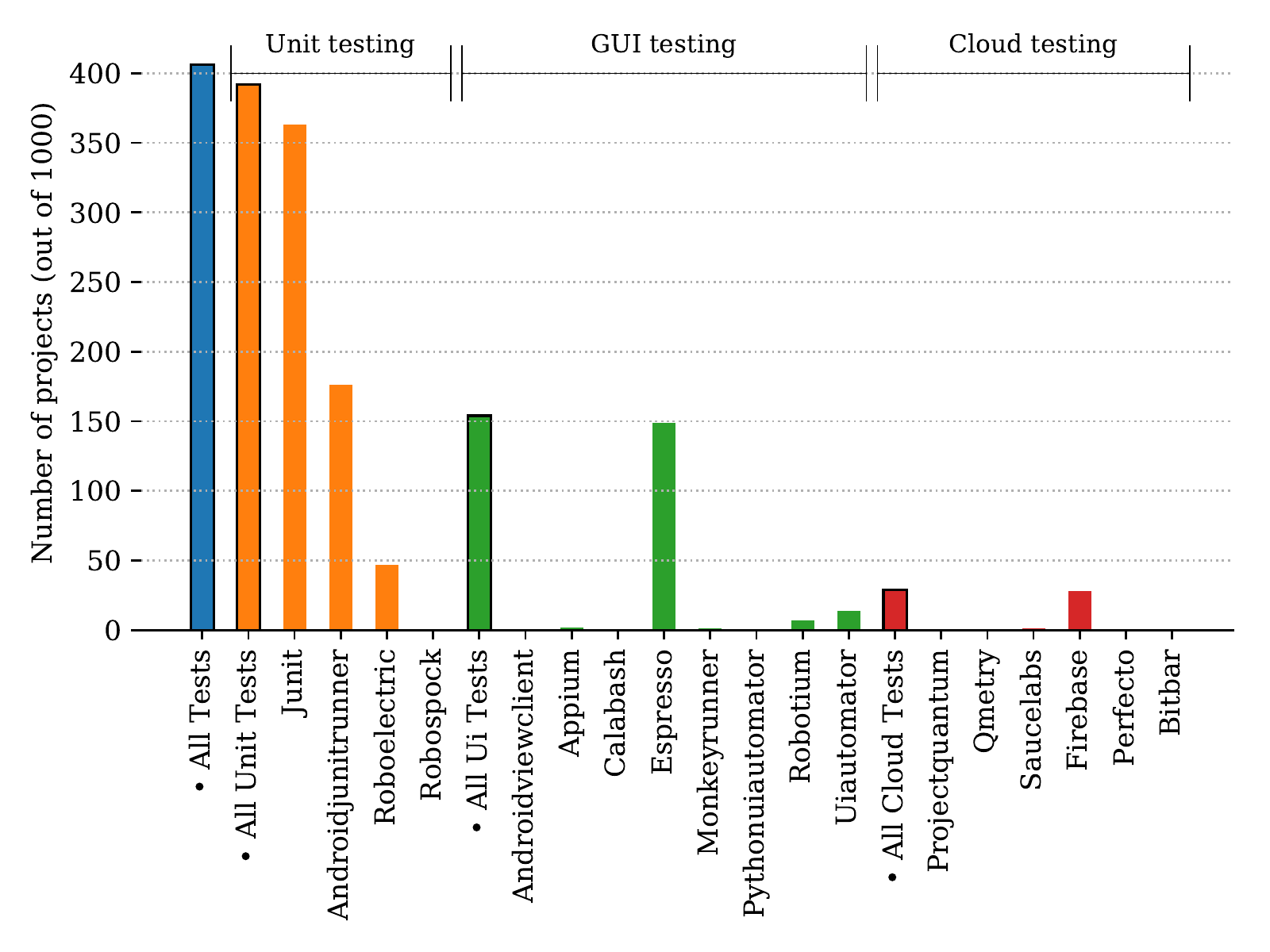}
  \caption{Number of projects per framework.}
  \label{fig:usage_count}
\end{figure}

Figure~\ref{fig:usage_count} shows, out of 1000 apps, the number of projects
using a test framework. We include results for \emph{Unit Testing}, \emph{UI
Testing}, and \emph{Cloud Testing} frameworks. The first bar shows the number
of apps that use any test tool. About $41\%$ of apps have tests. We can see
that unit tests are present in $39\%$ of projects while \emph{JUnit} is the
most popular tool, with $36\%$ of projects adopting it. This means that $89\%$
of projects with automated tests are using \emph{JUnit}.

Only $15\%$ of projects have automated User Interface (UI) tests.
\emph{Espresso} is the most used framework --- almost every project with UI tests is
using \emph{Espresso}. \emph{UIAutomator}, \emph{Robotium}, and \emph{Appium} are used by a very
small portion of projects in our dataset, while \emph{AndroidViewClient}, \emph{Calabash}, \emph{Monkeyrunner}, and
\emph{PythonUIAutomator} are not used in any project.

With less than $3\%$ of projects employing them, cloud testing services have
not found their way into the open source mobile app development community. In total, 28 projects use
\emph{Google Firebase}, whereas only 1 project uses \emph{Saucelabs}. All the
other cloud test services in this study are yet to be adopted.

\subsection{Discussion}

Most mobile apps published in \emph{F-droid} do not have automated tests. Developers are relying on
manual testing to ensure proper functioning of their apps, which is known to be less reliable and
to increase technical debt~\citep{stolberg2009enabling,bavani2012distributed,karvonen2017systematic}.

Given their simplicity, unit tests are the
most common form of tests. \emph{JUnit} is the main unit testing tool and the reason lies in the
official Android Developer documentation for tests\footnote{\emph{Getting Started with Testing}
Android guide available at: \url{https://goo.gl/RxmHq2}  (Visited on \today).}, which introduces \emph{JUnit} as the
basis for tests in Android. Furthermore, other test tools often rely internally on \emph{JUnit}
(e.g., \emph{AndroidJunitRunner}).

Other unit testing tools such as \emph{AndroidJunitRunner} and
\emph{Roboelectric} do not have a substantial adoption. These tools help running
unit tests within an Android environment, instead of the desktop's JVM. This is
important given the complexity of an Android app's lifecycle, which might
affect test results. However, many apps still do not cross that limit,
providing only unit tests for parts of the software that can run absent from
the mobile system. Since many apps follow a similar structure, based on Android's
framework enforced design patterns, easily customizable boilerplate tests
should be delivered along with those patterns.

UI tests are not so popular ($15\%$), which can be explained by their
cumbersome maintainability reported in previous
work~\citep{gao2016sitar,coppola2017fragility,li2017atom}. Although there are
many UI testing frameworks available, \emph{Espresso} is the only one with
substantial adoption. This is consistent with the phenomenon of \emph{JUnit}
for unit tests: \emph{Espresso} is also promoted in the official Android
Developer documentation. In fact, it is distributed with the Android Software
Development Kit (SDK). Another strength is that \emph{Espresso} provides
mechanisms to prevent flakiness and to simplify the creation and maintenance of
tests.

Previous work has considered \emph{Espresso} as the most energy efficient GUI testing
framework. The fact that these projects are already using it leaves an open door for the creation
of energy tests. On the other hand, \emph{Espresso} still provides a limited set of user
interactions, which can be a barrier to high test coverage~\citep{cruz2018measuring}.

Unfortunately, studied cloud testing services have not reached the open source
app community. This is probably due to the recency of the introduction of these
technologies and the lack of a testing culture in mobile app development, as
shown in our results.

The good news is that we observe an increasing adoption of unit and UI tests in
the last two years. This trend can be observed by comparing our findings with
previous work~\citep{kochhar2015understanding}; while the previous study
highlights that the prevalence of automated tests in mobile apps was merely
14\%, in this work, we observe that 41\% of FOSS apps are developed with
automated testing tools.

These findings provide useful implications for the development of new testing
tools and techniques. Previous work has shown the importance of creating new
types of tests for mobile apps (e.g., energy tests, security
tests)~\citep{linares2017continuous,muccini2012software,wang2015mobile}. Our
results show the importance of simplifying the learning curve and the project's
setup. Hence, new types of tests should be compatible at least with
\emph{JUnit} and \emph{Espresso}, avoiding reinventing the wheel or
complicating usage with new dependencies.

In addition, the adoption of these tools by the FOSS community is highly
sensitive to the quality and accessibility of documentation. The fact that
\emph{Google} has control over the official documentation does not help
third-parties to come aboard. Perhaps the official documentation should feature
more tools that are not delivered with the Android SDK.
The same concern applies to the academia that is developing many interesting
tools for mobile development and testing. Often the lack of documentation is a
big barrier to the adoption of innovative techniques by the software
industry~\citep{gousios2016work,kochhar2015understanding}.

\highlight{Only $41\%$ of FOSS apps have automated tests. Unit testing
frameworks are the most popular, comprising $39\%$ of projects. GUI testing is
being used by $15\%$ of projects, while the adoption of Cloud testing platforms
is negligible ($3\%$).}

\section{\rqthree{} (RQ2)} 
\label{sec:rqthree}


Android testing tools are in constant evolution to fit the ever-changing
constraints and requirements of mobile apps. Although we are currently far from
having a satisfactory prevalence of automated testing, the evolution from past
years can provide actionable information. We study which technologies and types
of testing have gained momentum, and which ones are still failing to be
perceived as beneficial in FOSS mobile app projects.

Thus, we analyze how the adoption of automated testing relates to the age of an
app and the time of an app's last update. We dig further and study the adoption
of automated testing in mature FOSS apps by years since the last update. Trends
on automated testing adoption over time are analyzed using scatter plots.

\subsection{Results}

\begin{figure}[htbp]
  \centering
    \includegraphics[width=0.7\columnwidth]{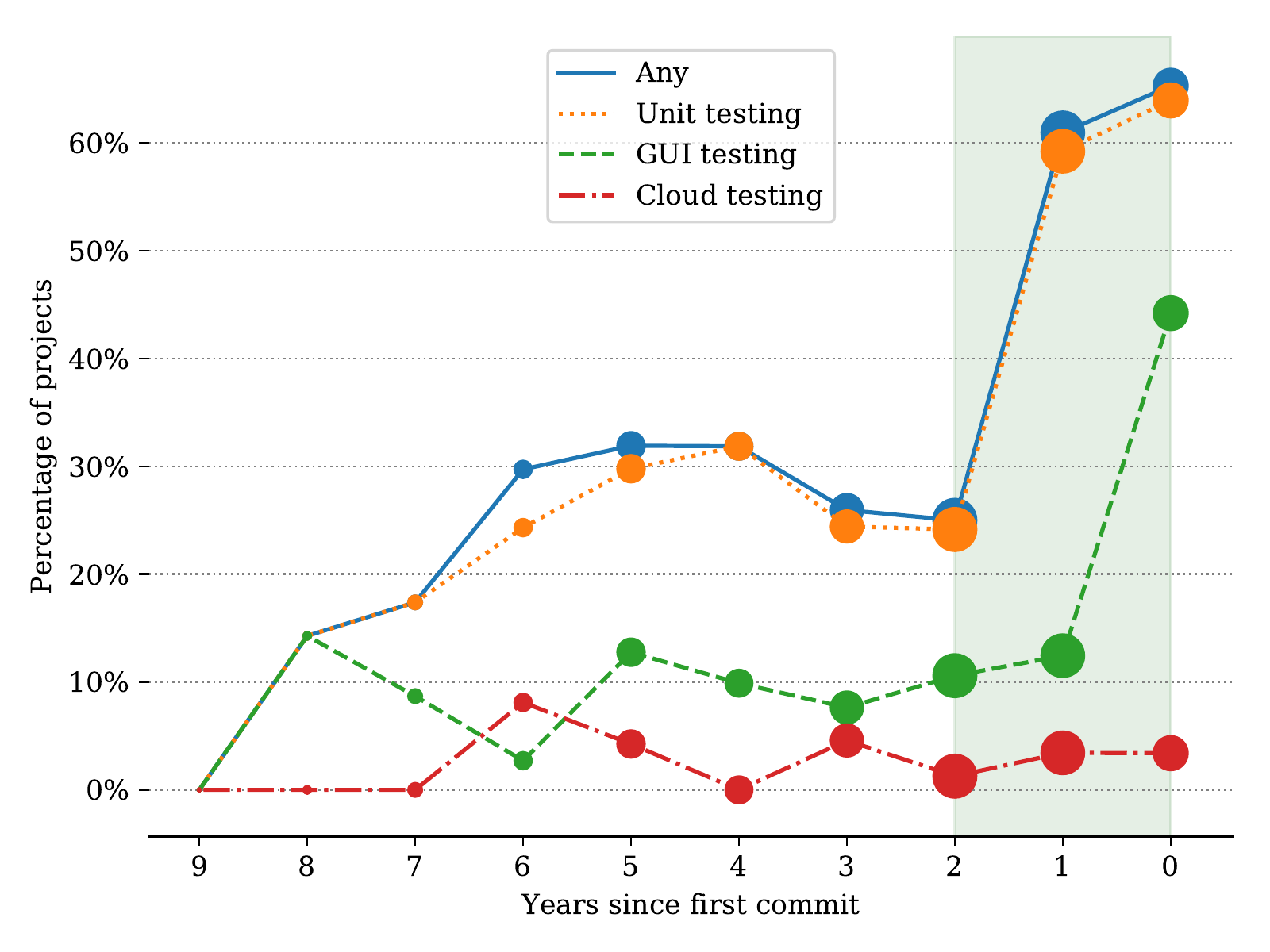}
  \caption{Percentage of Android apps developed in projects with test cases over the age of the apps.}
  \label{fig:tests_by_age}
\end{figure}

The percentage of apps that are doing tests grouped by their age is presented
in the plot of Figure~\ref{fig:tests_by_age}. The data is presented from older
to newer projects (i.e., 9--0 years old). The size of each circle is
proportional to the number of apps with that age (e.g., older projects have
smaller circles, meaning that there are fewer projects for those ages.). It is
used to show the impact of results in each case. E.g., since projects that are
six or more years old have small circles, they comprise a small number of
projects. Hence, trends in those age groups are not significant.


The timeline in Figure~\ref{fig:tests_by_age} shows that apps that are less
than two years old have significantly more tests than older apps. Moreover, the
usage of GUI testing frameworks has increased among apps that are under two
years old.

\begin{figure}[htbp]
  \centering
    \includegraphics[width=0.7\columnwidth]{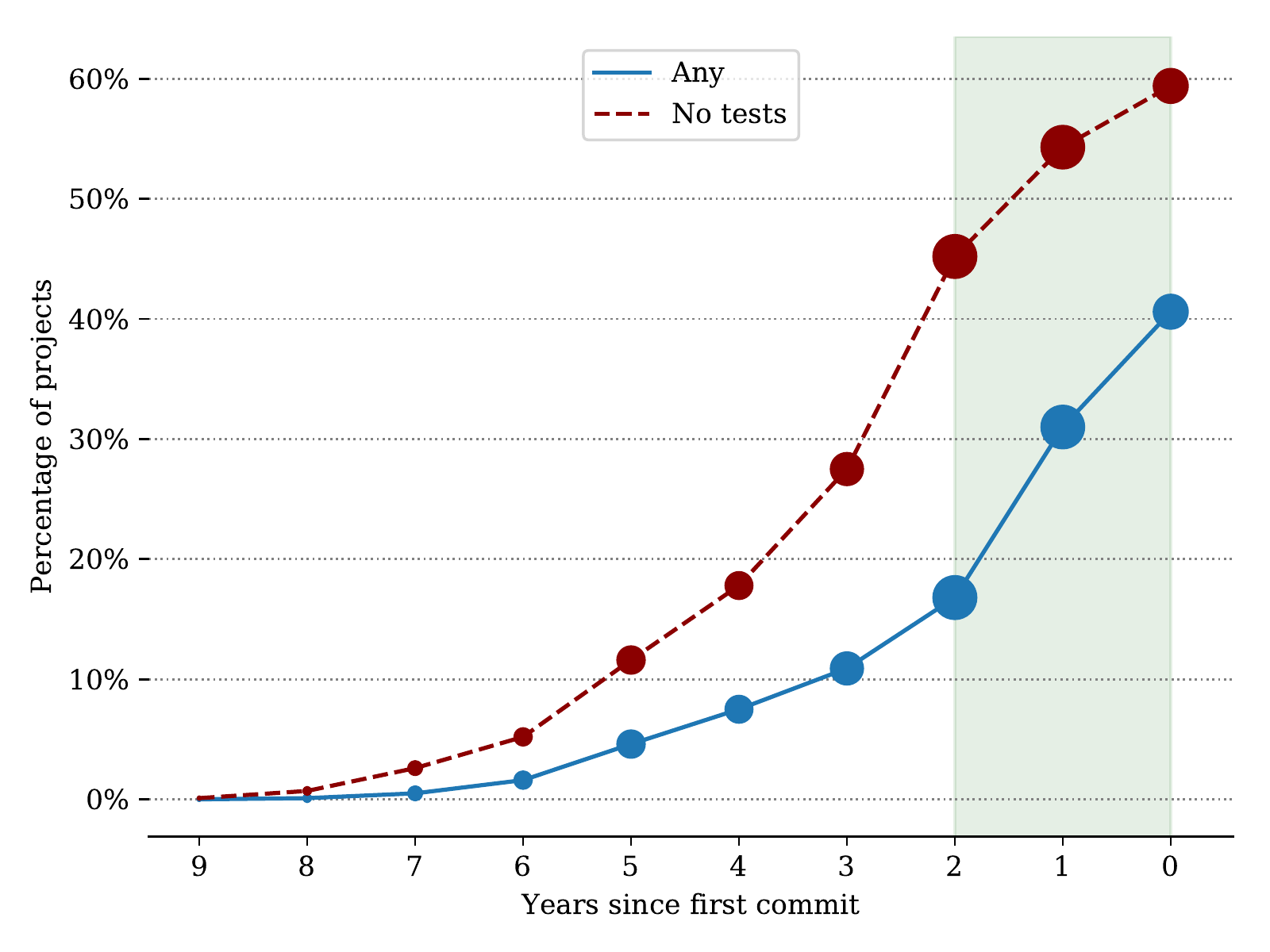}
  \caption{Cumulated frequency of projects with and without tests (from 9 to 0 years old),
 normalized by the total number of projects.}
  \label{fig:cum_tests_by_age}
\end{figure}

In addition, we present in Figure~\ref{fig:cum_tests_by_age} how new apps have
been changing the overall test automation adoption. In the past two years
(shaded region) the slope of projects with tests is higher than projects
without tests. However, this recent change is not able to change the overall
pervasion of test automation: most projects are not doing it.

\begin{figure}[htpb]
  \centering
    \includegraphics[width=0.7\textwidth]{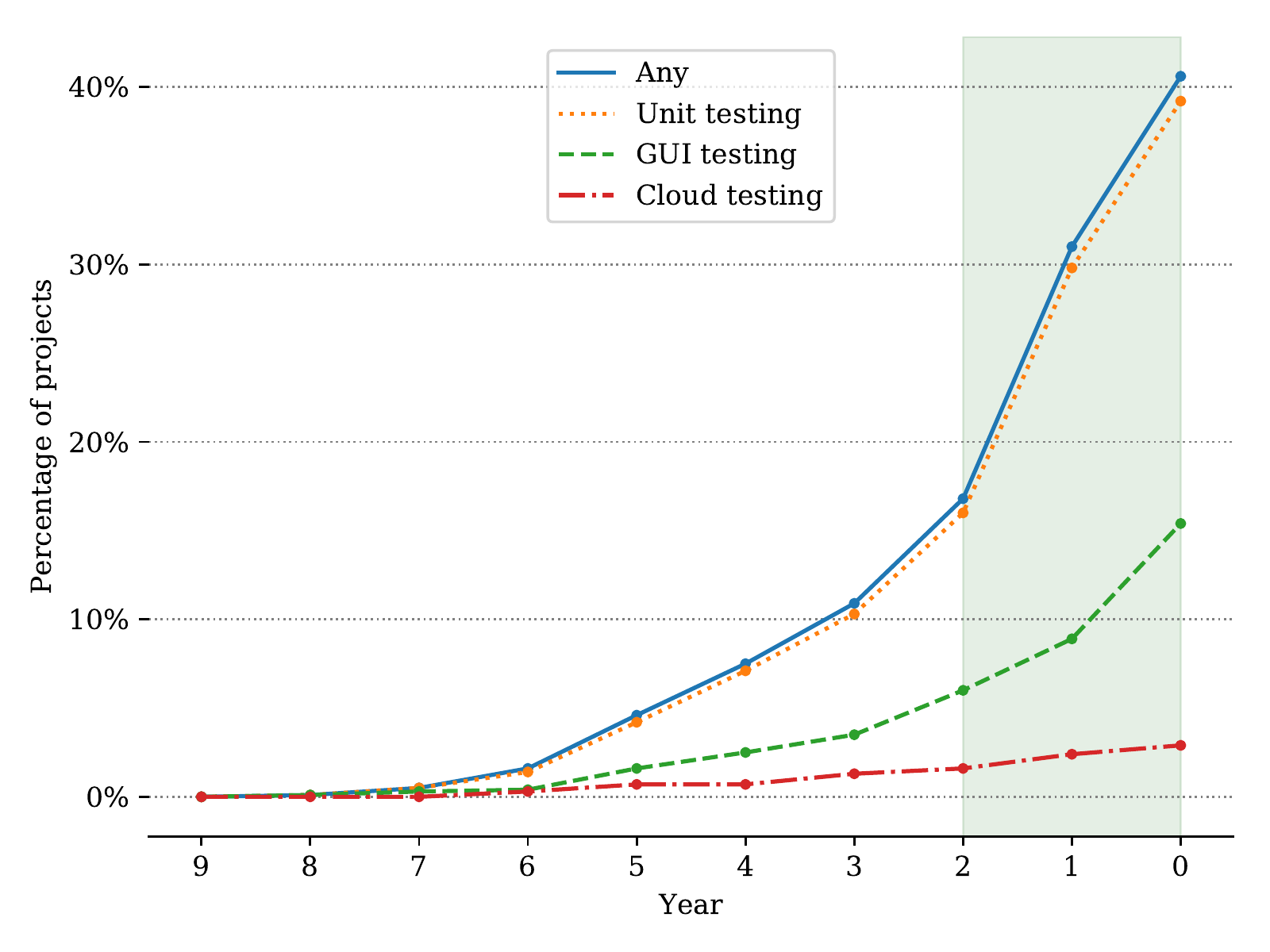}
  \caption{Cumulated percentage of projects with tests (from 9 to 0 years old),
 normalized by the total number of projects. All test categories are represented.}
  \label{fig:cum_tests_by_age_all}
\end{figure}

Finally, we present the timeline of the adoption for different kinds of testing
techniques in Figure~\ref{fig:cum_tests_by_age_all}. The aforementioned trend
is observable for unit testing and GUI testing, which have a higher slope in the
past two years (shaded region).

%
%

\subsection{Discussion}

Results show a significant increase in automated testing amongst new FOSS apps.
However, the fact that older apps have a lower adoption rate of automated
testing can be explained by two phenomena: 1) automated testing is becoming
more accessible to developers, who are becoming more aware of its benefits, 2)
at some point during the lifespan of a project, developers realize that the
overhead of maintaining automated testing is not worth the benefits and decide
to remove it. While the first phenomenon reveals a positive trend, the latter
is quite alarming --- automated testing does not provide a long-term solution.

Giving a better sense of which phenomenon is more likely to happen,
Figure~\ref{fig:cum_tests_by_age} reveals that automated testing has been
gaining popularity in the last two years.

It is worth noting that this increase is happening in both unit testing and GUI
testing. The fact that GUI testing is gaining popularity is important --- unit
testing per se does not provide means to achieve high test coverage in mobile
apps. This increase provides more case studies for researchers to study new
types of mobile testing (e.g., energy, security, etc.).

\highlight{Open source mobile developers are becoming more aware of the
importance of using automated tests in their apps. This is observed more for
apps that are updated recently than those updated several years ago.}

\section{\rqfour{} (RQ3)} 
\label{sec:rqfour}


In this study, we compare popularity metrics with the adoption of automated
testing practices in FOSS Android apps. The following popularity metrics were
selected:

\begin{itemize}

  \item []\textbf{Number of Stars.} The number of Github users that have marked the
  project as favorite.

  \item []\textbf{Number of Forks.} The number of Github users that have created a
  fork of the repository.

  \item []\textbf{Number of Contributors.} The number of developers that have contributed to the project.

  \item []\textbf{Number of Commits.} The number of commits in the repository.

  \item []\textbf{Average Rating.} The average user rating from \emph{Google Play} store.

  \item []\textbf{Number of Ratings.} The number of users rated the app on \emph{Google Play}.

\end{itemize}

These metrics depend on a myriad of factors, which do not necessarily relate to
mobile app development processes. Yet, they are notable metrics that developers
do care about. Typically, developers need to drive their development process
based on multiple sources of feedback~\citep{nayebi2018app}. We want to
investigate whether there is any kind of relationship between these features
and automated testing. Relationships can help motivate mobile app developers
employing tests in their projects.

To remove atypical cases, we perform an outlier detection using the Z-score
method with a threshold of three standard deviations. In addition, we perform
the normality test \emph{Shapiro-Wilk}, which tests the null hypothesis that
data follows a normal distribution.

Then we apply hypothesis testing, using the non-parametric test Mann-Whitney U,
with a significance level ($\alpha$) of $0.05$. We may also consider a
parametric test (e.g., the standard t-test), in case we find variables that
follow a Normal distribution. In addition, since we are conducting multiple
comparisons, the Benjamini-Hochberg procedure is used to correct $p$-values and
control false discovery rate.

The independent variable is whether an app has tests in its project source code
while the dependent variables are the popularity metrics.

The hypothesis test is formulated as follows, with populations $WO$ and $W$ as
the population of \textbf{apps without tests} and the population of
\textbf{apps with tests}, respectively:

\begin{equation*}
H_0: P(W>WO) = P(WO>W)
\end{equation*}
\begin{equation*}
H_1: P(W>WO) \neq P(WO>W)
\end{equation*}

In other words, we test the null hypothesis ($H_0$) that a randomly selected
value from population $W$ is equally likely to be less than or greater than a
randomly selected value from sample $WO$.

We perform hypothesis testing for
each of the aforementioned metrics, formulated as follows:

\begin{itemize}

  \hypothesis{Number of Stars}{a project with tests ($W$) has the same number
  of Github stars as a project without tests ($WO$)}{the number of Github Stars
  in projects with tests is different from the number of stars in a project
  without tests}

  \hypothesis{Number of Forks}{a project with tests ($W$) has the same number
  of forks as a project without tests ($WO$)}{the number of forks in projects
  with tests is different from the number of stars in a project without tests}

  \hypothesis{Number of Contributors}{projects with tests ($W$) have the same
  number of contributors as a project without tests ($WO$)}{the number of forks in
  projects with tests is different from the number of contributors in a project
  without tests}

  \hypothesis{Number of Commits}{a project with tests ($W$) has the same number
  of commits as a project without tests ($WO$)}{the number of commits in
  projects with tests is different from the number of commits in a project
  without tests}

  \hypothesis{Average Rating}{a project with tests ($W$) has the same rating as
  a project without tests ($WO$)}{the rating of a randomly selected project
  with tests is different from the rating in a project without tests}

  \hypothesis{Number of Ratings}{a project with tests ($W$) has the same number
  of rating as a project without tests ($WO$)}{the number of ratings of a
  randomly selected project with tests is different from the number of ratings
  in a project without tests}

\end{itemize}

In addition, we perform effect size analyses for variables showing statistical
significance. We compute the mean difference ($\Delta\bar{x} = \bar{x}_{W} -
\bar{x}_{WO}$), the difference of the medians ($\Delta Md = Md_{W} - Md_{WO}$),
and the Common Language Effect Size (CL)~\citep{mcgraw1992common}.

The mean difference ($\Delta\bar{x}$) measures the difference between the means
of apps with tests ($W$) and apps without tests ($WO$) for a particular
popularity metric. We compute it for being a conventional effect-size metric.
In addition, since the distribution is not necessarily normal, we compute the
difference of the medians ($\Delta Md$) between apps with tests ($W$) and apps
without tests ($WO$). Given that the median of a sample is the value that
separates the higher half from the lower half of the sample, $\Delta Md$
measures how different this median value is in the two distributions.

There are nonetheless a few cases in which $\Delta Md$ does not capture
differences in the two distributions~\citep{dave2014simple}. We complement the
effect size analysis with the Common Language (CL) measure. CL is the
recommended measure when there is no assumption on the shape of the
distributions of the two samples being tested and it is commonly used in tandem
with Mann-Whitney U test~\citep{leech2002call}. One advantage of using CL to
measure effect size is that it can be easily
interpreted~\citep{brooks2014common}: it is the probability that the value from
an individual randomly extracted from one sample will be higher than the value
from an individual randomly extracted from another.

\subsection{Results}

\begin{figure}[htpb]
  \centering
    \includegraphics[width=0.8\textwidth]{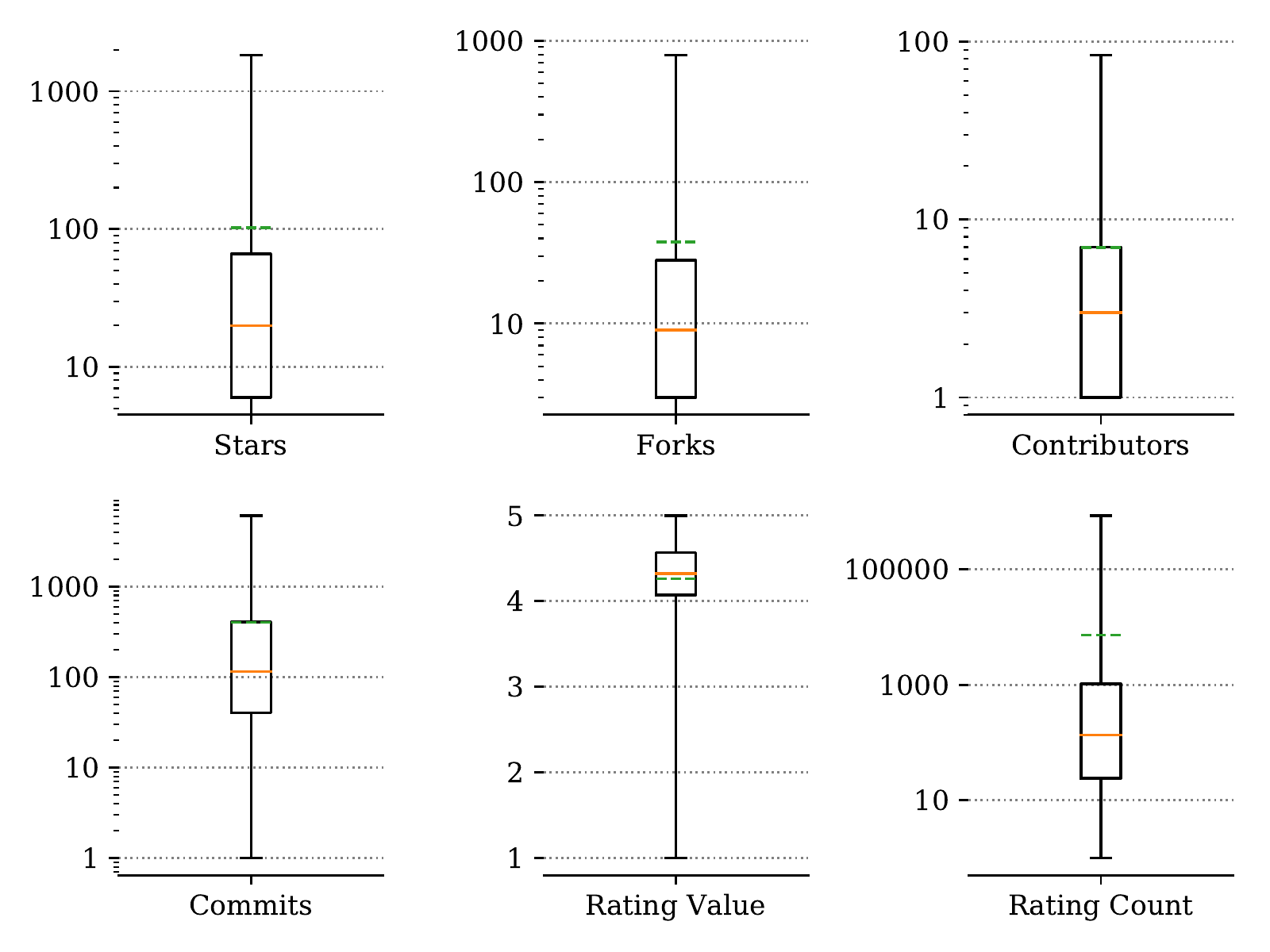}
  \caption{Boxplots with the distributions of the popularity metrics. Note
  that the y-axis is in log-scale for all metrics but ratings.}
  \label{fig:popularity_metrics}
\end{figure}

The distributions of the popularity metrics are depicted in the boxplots of
Figure~\ref{fig:popularity_metrics}. The medians are represented by the orange
solid lines, while the means are by green dashed lines. The results of the
normality tests \emph{Shapiro-Wilk} yielded a low $p$-value ($p<0.001$) for all
metrics. Thus, none of the metrics follows a normal distribution, which highlights
the suitability of using the Mann-Whitney U test over the standard t-test.

Hypothesis testing results are shown in Table~\ref{tab:popularity_metrics_test}
along with the effect size analysis: mean difference ($\Delta\bar{x}$),
difference of median ($\Delta Md$), and CL expressed in percentage. The
bigger the effect size is, the bigger is the metric for apps with tests. The
effect size analysis is only relevant in cases with statistical significance,
which are highlighted in bold text.

\begin{table}[htbp]
  \caption{Statistical analysis of the impact of tests on the popularity of apps.}
\begin{center}
  \label{tab:popularity_metrics_test}
\begin{tabular}{lrrrrc}
\hline
                          & $p$-value           & $\Delta\bar{x}$   & $\Delta Md$      & CL (\%)          \\
\hline                                                                                            
 Stars                    & $0.2130$            & $54.78$           & $3.00$           & $52.74\%$          \\
 Forks                    & $0.4525$            & $11.39$           & $1.00$           & $51.40\%$          \\
 \textbf{Contributors}    & $\mathbf{0.0052}$   & $\mathbf{2.17}$   & $\mathbf{0.00}$  & $\mathbf{55.80\%}$          \\
 \textbf{Commits}         & $\mathbf{0.0008}$   & $\mathbf{247.58}$ & $\mathbf{49.00}$ & $\mathbf{57.13\%}$          \\
 Rating Value             & $0.0835$            & $0.05$            & $0.05$           & $54.77\%$          \\
 Rating Count             & $0.2465$            & $-894.26$         & $-56.00$         & $47.03\%$          \\
\hline
\end{tabular}

\end{center}
\end{table}

There is statistical evidence that FOSS Android apps with tests are expected to
have more \textbf{commits} and more \textbf{contributors}. Note, however, that
this evidence does not imply that tests boost these variables. Conclusions must
analyze the causality of this relationship (i.e., whether tests are cause or
consequence) and the fact that there are many external variables that are
expected to have a significant impact (e.g., target users, originality of idea,
design, marketing, etc.). Nonetheless, no statistical significance was found
between having automated tests and the number of \emph{GitHub} stars, forks and
ratings on \emph{Google Play}.


Projects with tests have on average more $248$ commits in the whole project.
The CL effect size is small but substantial: the probability of a project with
tests having more commits than a project without tests is $57\%$. Although
the number of commits increases, one can argue that the number of commits can
be related to overhead created by tests maintenance.

Projects with tests have a small but substantial CL effect size: the
probability that a project with tests will have more contributors is $56\%$.
Nevertheless, the direction of this relationship cannot be assessed with these
results --- i.e., there is no evidence of whether the presence of tests is a
consequence of the high number of contributors in the project or, in contrary,
it is a way of attracting more developers to contribute.

\paragraph{Tests and Contributors: developers' perception?}

We decided to conduct a follow-up study to assess the developers' perception of
whether tests can lead to more contributors. We contacted 343 mobile open
source developers to answer a survey with two close-ended questions:

\begin{enumerate}

  \item \emph{Do you think that more tests benefit/attract new-comers?}\\
  Possible answers were: \emph{Yes}; \emph{No}; and \emph{Maybe}.

  \item \emph{Is the presence of tests a reason or a consequence of a big
  community of contributors?}\\ Possible answers were: \emph{Most likely a reason};
  \emph{Most likely a consequence}; \emph{Both equally}; and \emph{No impact}.

\end{enumerate}

Respondents had an additional box where they could optionally leave their
comments or feedback on the subject. Developers were selected by being active
in an open source mobile application available on \emph{GitHub}. In the end, we
had 44 responses. Data collected was anonymized and it is
available online\footnote{Questionnaire responses are available online:
\url{https://goo.gl/6CFDb9}}.

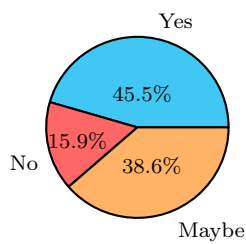
\begin{figure}[htbp]

  \centering
\begin{tikzpicture}[scale=0.4]
\pie [color={cyan!60,red!60,orange!60}]
    {45.5/Yes,15.9/No,38.6/Maybe}
\end{tikzpicture}
\caption{Do tests attract newcomers?}
\label{fig:response1}
\end{figure}

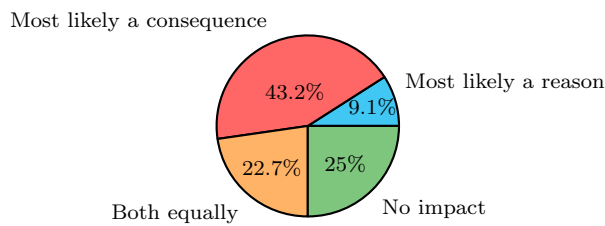
\begin{figure}[htbp]
  \centering
\begin{tikzpicture}[scale=0.4]
\pie [color={cyan!60,red!60,orange!60,darkgreen!60}]
    {9.1/Most likely a reason,43.2/Most likely a consequence,22.7/Both equally,25/No impact}
\end{tikzpicture}
\caption{Tests: cause or consequence of a big community?}
\label{fig:response2}
\end{figure}

As shown in the pie chart of Figure \ref{fig:response1}, 45.5\% of our respondents
believe that tests help new developers to contribute in a project, while 38.6\%
are not sure, and only 15.9\% disagree with the statement. The pie chart of
Figure \ref{fig:response2} shows that, despite the recognized improvement from having
tests, the majority of respondents believe that the presence of tests is more
likely a consequence from having a big community of contributors (43.2\%). A
smaller part of respondents (25\%) believe that the presence tests and the size
of the community do not affect each other --- i.e., they both depend on a
different variable. Other respondents believe both variables affect each other
(22.7\%), while only 9.1\% reckon tests as the cause.

Feedback submitted by some developers provided some insights on their
personal experience. Some developers pointed out that the adoption of CI/CD is
probably ``more influential than the actual tests''. Other developers
emphasized the importance of having tests as ``a good starting point for
newcomers to get familiar with the project's code and its features''. Finally,
some developers state that the ``maintenance burden of automated tests is
really high'' and that they can block major refactorings in software projects.

\subsection{Discussion}

Results show that FOSS Android projects with tests have more commits and more
contributors. The increase in the number of commits can be explained by an overhead
of commits induced by the maintenance and configuration of tests.

Responses to the questionnaire show that the presence of tests is more likely a
consequence of having a big community. In addition, tests can help new
developers contribute to the project. Since one of the main concerns in open
source projects is to foster the community to contribute\footnote{\emph{Five
best practices in open source: external engagement} by Ben Balter:
\url{https://goo.gl/BQRZBa} (Visited on \today).}, the importance of tests for
this purpose cannot be discarded. Conventionally, maintainers of open source
projects target this goal by inviting contributors, providing social and
communication tools, and making sure that instructions on how to contribute are
well documented. These results show that tests should also be part of their
agenda.

This relationship is consistent with previous work. Automated tests help new
developers be more confident about the quality of their
contributions~\citep{gousios2016work}. Contributors are able to create
\emph{Pull Requests} (PR) to a project with a reasonable level of confidence
that other parts of the software will not break. The same applies to the
process of validating a PR. Integrators usually have some barriers when
accepting contributions from newcomer developers~\citep{gousios2015work}. The
presence of automated tests helps reduce that barrier, and contributions with
tests are more likely to be accepted~\citep{gousios2015work}. Another aspect of
automated tests that contributes to this trend is the reported ability to
provide up-to-date documentation of the
software~\citep{van2001program,beck2000extreme}.

Previous work that shows that app store's ratings are not able to capture the
quality of apps~\citep{langhe2015navigating,ruiz2017examination}. Our results
show that this is also the case for tests: there is no relationship between
using tests and rating on \emph{Google Play}.

Our findings have direct implications for different stakeholders of mobile
software projects. Developers have to start using automated tests in their code
in order to assure quality in their contributions. Open source project
maintainers must promote a testing culture to engage the community in their
projects.

\highlight {Automated testing is important to foster the community to
contribute. There is statistical evidence that FOSS Android projects with tests
have more contributors and more commits. Number of \emph{GitHub Stars},
\emph{Github Forks} and ratings on \emph{Google Play} did not reveal any
significant impact.}

\section{\rqfive{} (RQ4)} 
\label{sec:rqfive}


Code issues are related to potential vulnerabilities of software. It is a
major concern of developers to ship software with a minimal number of code
issues. We study whether automated testing can help developers deploy mobile
app software with fewer code issues.

We use the issues detected by the static analysis Sonar tool as a proxy of
software code issues. We apply Sonar to our dataset of $1000$ Android apps. As
mentioned in Section~\ref{sec:data_collection}, Sonar issues are divided into
four categories, based on the severity of their impact. We evaluate the number of
issues normalized for the number of files in the project ($I'(p)$).

We apply the same approach used in Section~\ref{sec:rqfour}: we use hypothesis
testing with the Mann-Whitney U test using a significance level ($\alpha$) of
$0.05$. Benjamini-Hochberg procedure is used to correct $p$-values since four
tests are performed in the same sample. Mean difference ($\Delta\bar{x}$),
difference of median ($\Delta Md$), relative difference
($\frac{\Delta{}Md}{Md_W}$), and CL are used to analyze effect size.

\subsection{Results}

We successfully collected code issue reports from $967$ apps. It was not
possible to collect data from 33 apps: Sonar failed due to characters invalid
with UTF-8 encoding. This was the case of the reading app \emph{FBReaderJ} and
its file \texttt{ZLConfigReader.java} that contained characters that not even
Github is able to render\footnote{Example of a source code file incompatible
with Sonar tool: \url{https://git.io/fxNg9} (Visited on \today).}. Since these
apps consisted of a small portion of our dataset (3\%), we decided to leave
them out of this part of the study.

Table~\ref{tab:sonar_metrics} presents descriptive statistics of the number of
code issues per file $I'(p)$ for each level of severity --- size of the sample
($N$), median ($Md$), mean ($\bar{x}$), and standard deviation ($s$). The table
also presents the results of normality tests with the $p$-values for
Shapiro-Wilk tests ($X \sim N$), showing that none of the metrics follows a
normal distribution. Statistics are presented for both apps with tests ($W$)
and apps without tests ($WO$).

\begin{table}[htbp]
  \caption{Descriptive statistics of code issues on apps with ($W$) and without ($WO$) tests}
\begin{center}
  \label{tab:sonar_metrics}
\begin{tabular}{llrrrrr}
\hline
                           & Tests   &   $N$ & $Md$   & $\bar{x}$   & $s$    & $X \sim N$   \\
\hline
 \multirow{2}{*}{Blocker}  & $W$     &   398 & $0.00$ & $0.02$      & $0.04$ & $p < 0.0001$ \\
                           & $WO$    &   569 & $0.00$ & $0.05$      & $0.59$ & $p < 0.0001$ \\
\hline
 \multirow{2}{*}{Critical} & $W$     &   398 & $0.24$ & $0.34$      & $0.39$ & $p < 0.0001$ \\
                           & $WO$    &   569 & $0.26$ & $0.48$      & $0.90$ & $p < 0.0001$ \\
\hline
 \multirow{2}{*}{Major}    & $W$     &   398 & $0.50$ & $0.73$      & $0.80$ & $p < 0.0001$ \\
                           & $WO$    &   569 & $0.52$ & $0.84$      & $1.09$ & $p < 0.0001$ \\
\hline
 \multirow{2}{*}{Minor}    & $W$     &   398 & $0.61$ & $0.87$      & $0.93$ & $p < 0.0001$ \\
                           & $WO$    &   569 & $0.73$ & $1.27$      & $2.12$ & $p < 0.0001$ \\
\hline
\end{tabular}
\end{center}
\end{table}

\begin{figure}[htpb]
  \centering
    \includegraphics[width=0.8\columnwidth]{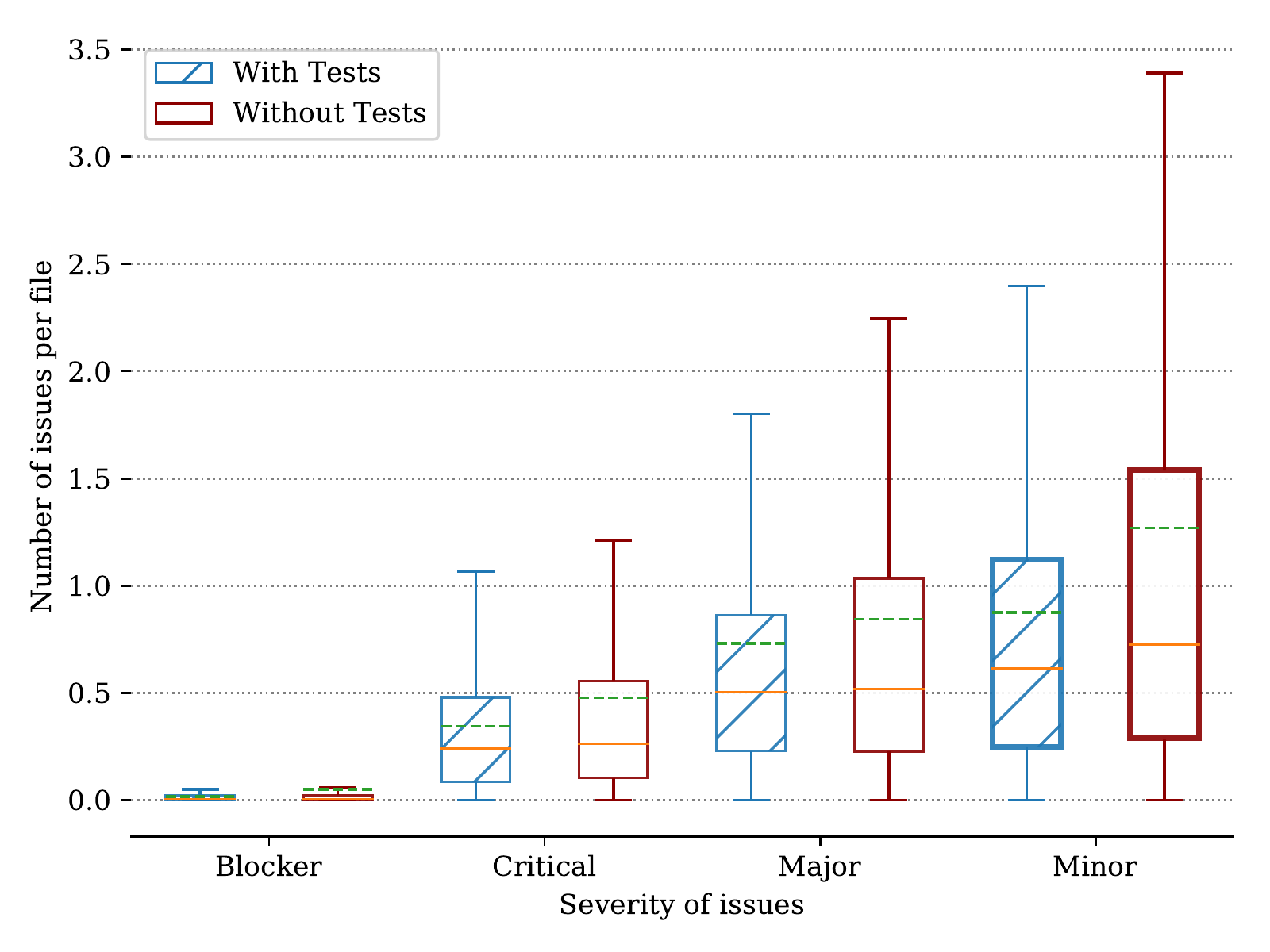}

  \caption{Comparison of the number of issues per file in projects with and without
  tests. Green dashed lines in each box represent the mean value, while orange
  solid lines represent the median.}

  \label{fig:sonar_vs_tests}
\end{figure}

Figure~\ref{fig:sonar_vs_tests} illustrates the distribution of the $I'(p)$ in
projects with tests (blue line, hatch fill) and without tests (red
line, empty fill) for different types of issues. The mean for each group is
depicted with a dashed green line, while the median with a solid orange line.
Types of issues with a statistically significant difference between W and WO
are highlighted with thicker lines. Results show that projects with tests have
significantly less minor code issues than projects without tests.

Table~\ref{tab:sonar_metrics_test} reports the resulting $p$-values and
computes the effect-size metrics: mean difference ($\Delta\bar{x}$), difference
of median ($\Delta Md$), relative difference ($\frac{\Delta{}Md}{Md_W}$), and
CL.

The number of minor issues per file increases significantly in projects without
tests. The difference of median shows that projects without tests are expected
to have 0.11 more minor issues per file (increase of 18\%). Furthermore, as
reported with the CL effect-size, projects without tests have more minor issues
than projects with tests with a probability of 54\%. The number of issues for
higher severity levels is not significantly affected.

\begin{table}[htbp]
  \caption{Statistical analysis of the impact of tests in mobile software code issues}
\begin{center}
  \label{tab:sonar_metrics_test}
\begin{tabular}{lrrrrr}
\hline
 Severity       &       $p$-value &   $\Delta\bar{{x}}$ &     $\Delta Md$ &   $\frac{\Delta{}Md}{Md_W}$(\%)& CL (\%)    \\
\hline                                                                                                        
 Blocker        &          0.1643 &              0.0337 &          0.0014 &                             48 & 52.09\%      \\
 Critical       &          0.1150 &              0.1337 &          0.0234 &                              9 & 52.97\%      \\
 Major          &          0.2939 &              0.1130 &          0.0157 &                              3 & 51.02\%      \\
 \textbf{Minor} & \textbf{0.0440} &     \textbf{0.3940} & \textbf{0.1127} &                    \textbf{18} & \textbf{54.32\%}      \\
\hline
\end{tabular}
\end{center}
\end{table}

\subsection{Discussion}

Results show that there is a statistically significant and substantial
relationship between using automated testing and the number of minor code
issues that appear in the project. FOSS Android projects without automated
testing have significantly more minor code issues. Given that only $41\%$ of
apps in this study have automated tests, mobile developers need to be aware of
the importance of testing their apps.

On the other hand, although the normalized number of blocker, critical,
and major bugs is higher for apps without tests than those with tests, the
difference is not statistically significant. Other alternatives, such as manual
testing, code inspection, or static analysis, are probably preventing such
issues. Our sample size may also not be large enough to make the result to be
statistically significant.

\highlight{There is statistical evidence that FOSS Android projects without
tests have $18\%$ more minor code issues per file. In our sample, projects
without tests also had more code issues for other severity levels: major
(3\%), critical (9\%), and blocker (48\%).}

\section{\rqsix{} (RQ5)}
\label{sec:rqsix}


CI/CD has been proved to be beneficial in software projects and to have even
better results when employed along with automated
testing~\cite{hilton2016usage,zhao2017impact}. Thus, we study whether mobile
app developers are using CI/CD in its full potential. Moreover, we delve into
how mobile app projects set themselves apart from conventional software
projects in terms of CI/CD adoption.

To answer this research question, we start by comparing the adoption of the
different studied CI/CD technologies in Android FOSS projects. In addition, we
compare the frequency of projects that have adopted one of the studied CI/CD
tools with the frequency of projects using automated testing.

For this analysis, we resort to data visualizations. To
validate the relationship between automated testing and CI/CD we use Pearson's
chi-squared test with a significance level of $0.05$. This test was selected for
being commonly used to compare binary variables.

\subsection{Results}
\begin{figure}[htpb]
  \centering
    \includegraphics[width=.6\columnwidth]{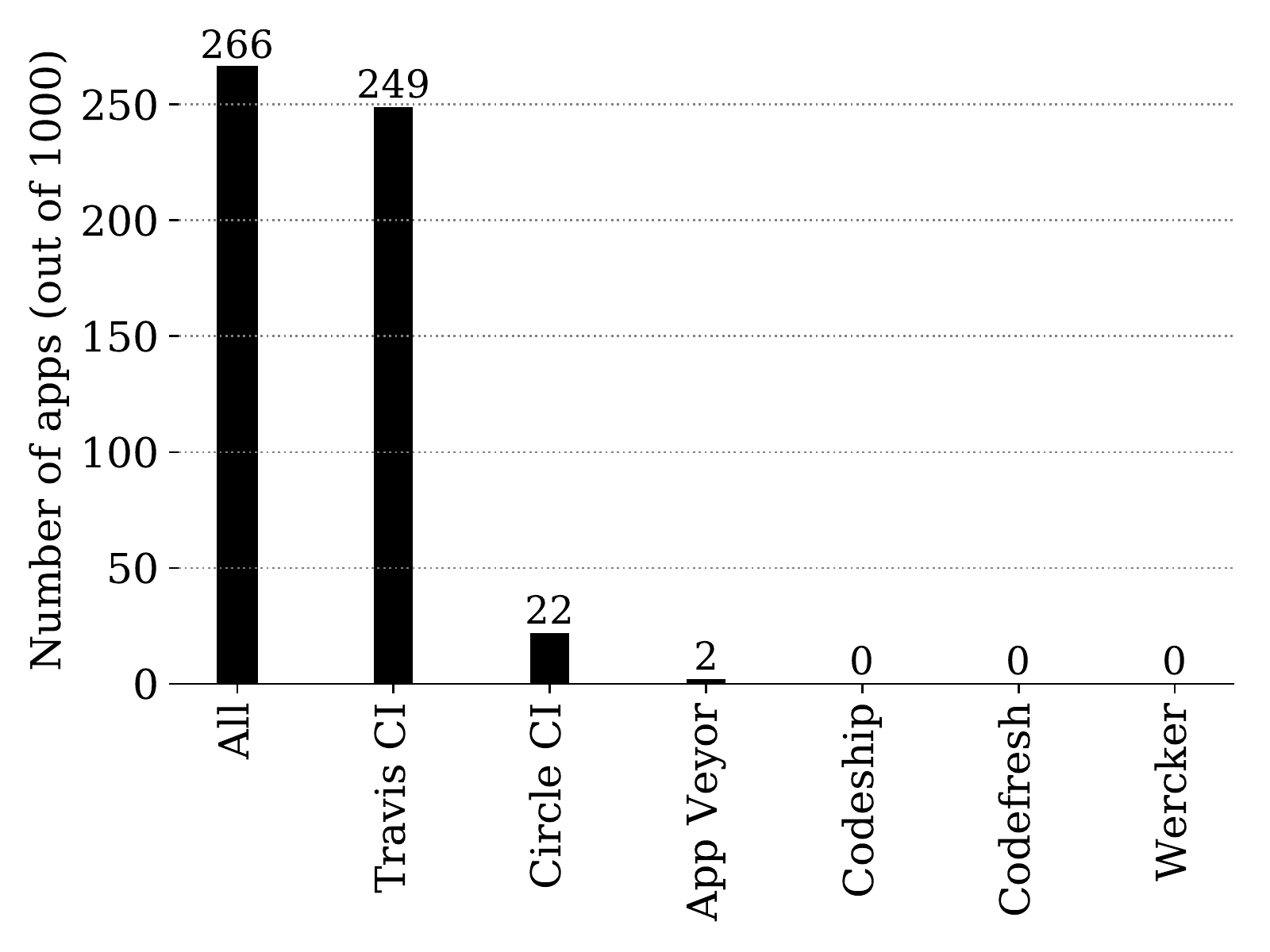}
  \caption{Android apps using CI/CD platforms.}
  \label{fig:reports_ci_cd_hist}
\end{figure}

We first analyze which apps are using CI/CD pipelines in their development practices.
The distribution of CI/CD pipelines among these platforms is given in Figure~\ref{fig:reports_ci_cd_hist}.
\emph{Travis CI} is the most popular platform with $249$ apps using it ($25\%$), followed by
\emph{Circle CI}, being used by $2\%$ of apps. However, in total, only $27\%$ have adopted
CI/CD.

\begin{figure}[htpb]
  \centering
    \includegraphics[width=0.7\columnwidth]{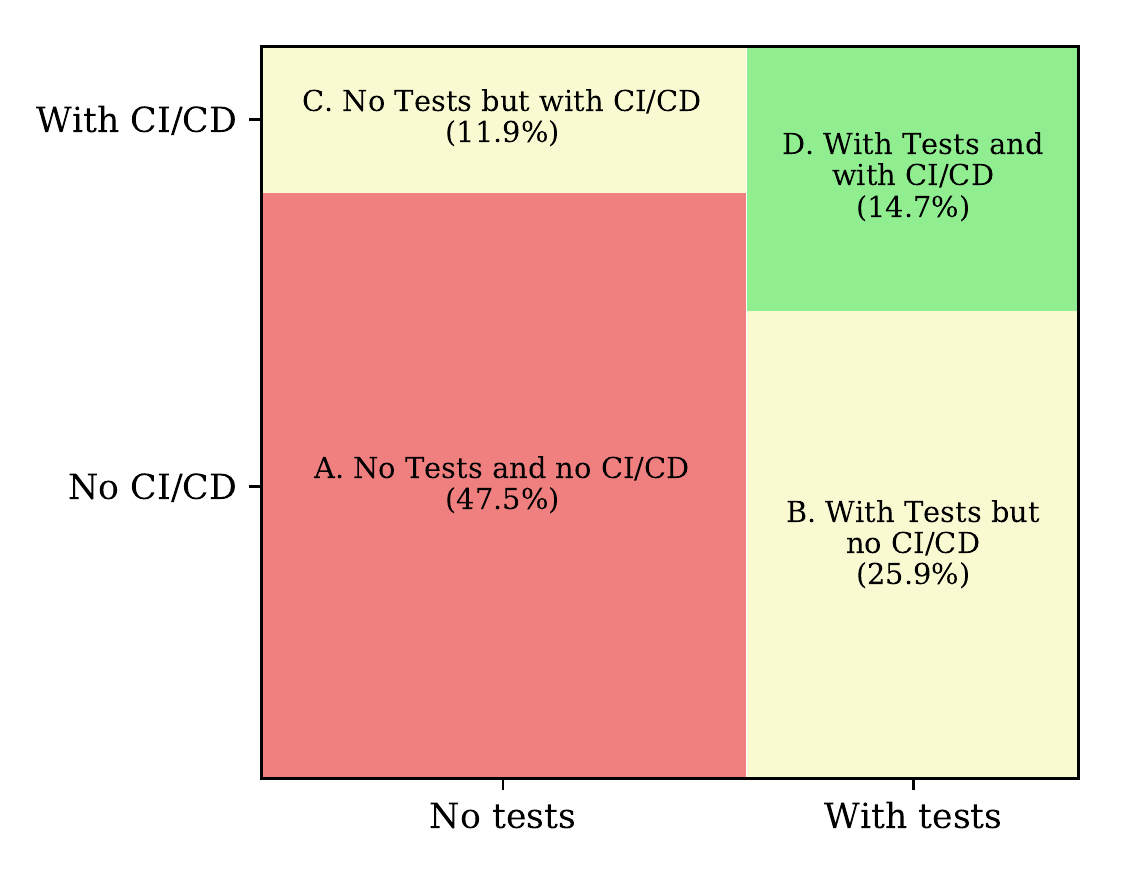}
  \caption{Relationship between apps using CI/CD and apps using tests.}
  \label{fig:reports_ci_cd_mosaic}
\end{figure}

The relationship between the prevalence of CI/CD and prevalence of tests is
depicted by the mosaic plot in Figure~\ref{fig:reports_ci_cd_mosaic}. The size
of each area is proportional to the number of apps in each group. Nearly $50\%$
of apps are not having tests nor adopting CI/CD (region A). $26\%$ of apps,
despite having tests, are not using CI/CD (region B). $12\%$ of apps are using
CI/CD but are not doing any automated tests (region C). Only $15\%$ of apps are
using CI/CD effectively, with automated tests (region D). In addition, the
mosaic plot suggests that automated testing is more prevalent in projects with
CI/CD than projects without. This is confirmed by the Pearson's chi-squared
test: $\chi^2 = 31.48, p = 2.009\text{e-}8$.

Online coverage trackers are useful tools that play well with CI/CD platforms.
They help ensure that the code is fully covered. Nevertheless, only $19$ projects
are using it --- $9$ use \emph{Coveralls} and $12$ use \emph{Codecov}, having
$2$ projects using both platforms. However, only $4$ have line coverage
above $80\%$, and no meaningful results can be extrapolated.

\subsection{Discussion}

CI/CD is not as widely adopted by mobile app developers as compared to
developers of general OSS projects --- only $26\%$ of apps have adopted CI/CD
services while the adoption in general open source software hosted by
\emph{GitHub} is $40\%$~\citep{hilton2016usage}.

There are $12\%$ of apps that, despite using CI/CD, do not have automated tests. In practice, these
projects are only using CI/CD tools to run static analyses. Yet, they rely on a pipeline
that requires an approver to manually build and test the app.

The fact that there are projects that have tests but did not adopt CI/CD
($26\%$) is also concerning. One of the main strengths of adopting CI/CD is
improving software quality through test automation~\citep{zhao2017impact}.
Although CI/CD services have made a good work in simplifying the configuration
of Android specific requirements (e.g., SDK version, emulator, dependencies,
etc.), developers have reported that the main obstacle in adopting CI/CD in a
project is having developers who are not familiar with
it~\citep{hilton2016usage}. Nevertheless, since these projects are already
using automated tests, they could potentially benefit from a CI/CD pipeline
with little effort. More research needs to be conducted to assess why mobile
developers are not adopting CI/CD in their projects.

\emph{Travis CI} and \emph{Circle CI} are the most used CI/CD services, as
expected from previous results for other types of
software~\citep{hilton2016usage}. Although the other platforms have a well
documented support for Android, they are not being used by the community.

Even more surprising is the fact that, from the $147$ apps with both CI/CD and
tests, only $19$ are actually promoting full test coverage with coverage
tracking services. This suggests that coverage is not a top priority metric for
mobile developers, which is in sync with concerns by Gao \textit{et al.} who
have reported the need for coverage criteria to meet the idiosyncrasies of
mobile app testing~\citep{gao2014mobile}. In particular, \emph{Coverall} and
\emph{Codecov} platforms only report line coverage. Different coverage
criteria, such as event/frame coverage, would be more suitable in the context
of mobile apps.

More education and training is needed to get full benefits of CI/CD for mobile
apps. Developers that are already performing automated tests in their apps
should explore the integration of a CI/CD pipeline in their projects. This is
also a good opportunity for newcomer developers willing to start contributing
to open source projects.

\highlight{CI/CD in mobile app development is not as prevalent as in other platforms;
Automated testing is more prevalent in projects with CI/CD.}

\section{Hall of Fame}
\label{sec:hof}

We have selected a set of apps from our dataset that we consider good
candidates for studying best practices from the mobile app development community.
We perform a systematic selection by choosing projects that perform unit tests,
UI tests and are using CI/CD. In total, 54 apps satisfy these
requirements\footnote{The whole set of apps in the Hall of Fame can be accessed
online: \url{https://luiscruz.github.io/android_test_inspector/}.}. We present
in Table~\ref{tab:hall_of_fame} one app for each category based on the
popularity of that app among developers, using the number \emph{GitHub Stars}
as a proxy. Some categories, namely \emph{Games}, \emph{Money}, and \emph{Phone
\& SMS}, did not have any app that meets the requirements.

\begin{table}[htbp]
  \caption{Hall of fame}
\begin{center}
  \label{tab:hall_of_fame}
  \begin{tabular}{lll}
\hline
 Category            & Organization   & Project Name            \\
\hline
 Internet            & k9mail         & k-9                     \\
 Multimedia          & TeamNewPipe    & NewPipe                 \\
 Writing             & federicoiosue  & Omni-Notes              \\
 Theming             & Neamar         & KISS                    \\
 Time                & fossasia       & open-event-android      \\
 Sports \& Health     & Glucosio       & android                 \\
 Navigation          & grote          & Transportr              \\
 System              & d4rken         & reddit-android-appstore \\
 Reading             & raulhaag       & MiMangaNu               \\
 Security            & 0xbb           & otp-authenticator       \\
 Science \& Education & EvanRespaut    & Equate                  \\
 Connectivity        & genonbeta      & TrebleShot              \\
 Development         & Adonai         & Man-Man                 \\
 Graphics            & jiikuy         & velocitycalculator      \\
\hline
\end{tabular}
\end{center}
\end{table}

Note, however, that although these projects follow best practices, they are not
necessarily the ones with the highest ratings (e.g., rating in \emph{Google Play},
number of Forks in \emph{Github}). The success of apps also depends on a myriad
of other factors. Nevertheless, the impact of best practices is not negligible
and for that reason, these projects can be used as role models for new projects
or subjects for case studies for further research.

%
%
%
%
%
%
%

\section{Threats to validity}
\label{sec:t2v}

\paragraph{Construct validity}

Code issues collected with \emph{SonarQube} are used to measure the quality of
code. Some projects might not follow common development guidelines due to
specific requirements. Thus, generic static rules might not be able to capture
the quality of such projects. Nevertheless, we expect that this is the case of
a minimal number of apps and results are not affected. Metrics from \emph{Google
Play} and \emph{GitHub} are used as proxies to measure user satisfaction, and
popularity of apps. These metrics are affected by a number of factors and not
always are sufficiently dynamic to cope with changes in the
app~\citep{ruiz2017examination}.

Furthermore, the online coverage trackers investigated in this study only
support line coverage. Coverage metrics for events or UI frames are more
suitable for mobile applications. These metrics were not evaluated as they are
not available in the state-of-the-art online coverage trackers. Finally, we did
not consider AIG techniques since they are more advanced and thus are not
popularly used in mobile app development yet.

\paragraph{Internal validity}

The usage of a test framework or service was assessed through a self-developed
automatic tool based on static analysis and Web requests to service's APIs. To
validate the accuracy of our tool we have manually labeled a random sample of
50 apps and compared the results. Our tool has successfully passed our
validation with no false positives and no false negatives, but we understand
that some corner cases may not have been checked yet. The same applies to the
static analysis tool \emph{SonarQube} used to collect code issues --- it
provides an approximation of the actual set of code issues in a project. Some
issues detected by \emph{SonarQube} may be false positives or may not
generalize to other, distinct projects. Nevertheless, we argue such cases are
rare and they are not expected to have a significant effect in results.

\paragraph{External validity}

Our work has focused on free and open source apps. Our 1000-app dataset
comprises a good proportion of these apps that are currently available for
Android users. Findings in this work are likely to generalize to types of apps
with a caveat: private companies usually have a different approach from open
source organizations on software testing~\citep{joorabchi2013real}. We did not
include testing services without a free plan for open source projects; Paid
apps have different budgets and might be more willing to use paid services in
their projects. Legal and copyright restrictions do not allow us to scope apps
with commercial licenses. This is a known barrier for research based on app
store analysis~\citep{krutz2015dataset,nagappan2016future,martin2017survey}.

The adoption of CI/CD is based on a subset of CI/CD services available, as described in
Section~\ref{sec:data_collection}. This subset is equivalent to the one used by
~\citeauthor{hilton2016usage} to study CI/CD adoption in general software projects~\citep{hilton2016usage}.

%
%
%
%
%
%
%
%
%
%
%

\section{Conclusion}
\label{sec:conclusions}
Testing is a crucial activity during the software development lifecycle to
ascertain the delivery of high quality (mobile) software. This study is about
testing practices in the mobile development world. In particular, we
investigated working habits and challenges of mobile app developers with
respect to testing.

A key finding of our large-scale study, using 1000 Android apps, is that mobile
apps are still tested in a very \emph{ad hoc} way, if tested at all. We show
that, as in other types of software, testing increases the quality of apps
(demonstrated in the number of code issues). The adoption of tests has increased
over the last two years and that Espresso and \emph{JUnit} are the most popular
frameworks. Furthermore, we find that availability of tests plays a positive
role in engaging the community to contribute to open source mobile app
projects. Yet another relevant finding of our study is that CI/CD pipelines are
rare in the mobile app world (only 26\% of the apps are developed in projects
leveraging CI/CD) -- we argue that one of the main reasons is due to the lack
of a good set of test cases and adoption of automatic testing. We have
discussed possible reasons behind the observed phenomena and some implications
for practitioners and researchers.

As future work, our empirical study can be expanded in several ways: 1) study
how mobile app projects address tests for particular types of requirements
(e.g., security, privacy, energy efficiency, etc.); 2) based on the test
practices collected from mobile app repositories, provide a set of best
practices to serve as rule of thumb for other developers; and 3) verify
that these findings also hold for other platforms.


\balance
\bibliographystyle{IEEEtranN}
\bibliography{bibliography}
\end{document}